\documentclass[twocolumn]{aastex631}

\usepackage{units}
\usepackage{mathtools}
\usepackage{etoolbox}
\defcitealias{Duffell+Laskar+2018}{DL18}
\defcitealias{Granot+Piran+2012}{GP12}
\defcitealias{Rhoads+1999}{R99}

\makeatletter
\patchcmd{\ltx@foottext}{%
  .5\textwidth\advance\hsize-18pt}{%
  \linewidth\advance\hsize-1.8em%
}{}{}
\makeatother

\begin{document}

\title{Did Binary Neutron Star Merger GW170817 Leave Behind A Long-lived Neutron Star?}

\correspondingauthor{Marcus DuPont}
\email{md4469@nyu.edu}

\author[0000-0003-3356-880X]{Marcus DuPont}
\author[0000-0002-0106-9013]{Andrew MacFadyen}
\affiliation{Center for Cosmology and Particle Physics, New York University \\
New York, NY, 10003, USA}



\begin{abstract}
We consider the observational implications of the binary neutron star (BNS) merger GW170817 leaving behind a rapidly rotating massive neutron star that launches a relativistic, equatorial outflow  as well as a jet. We show that if the equatorial outflow (ring) is highly beamed in the equatorial plane, its luminosity can be ``hidden'' from view until late times, even if carrying a significant fraction of the spin-down energy of the merger remnant. 
This hidden ring reveals itself  as a re-brightening in the light curve 
once it slows down enough for Earth to be within the ring's relativistic beaming solid angle. We compute semi-analytic light curves using this model and find they are in agreement with the observations thus far, and we provide predictions for the ensuing afterglow.
\end{abstract}



\section{Introduction} \label{sec:intro}
The historic gravitational wave (GW) event, GW170817 \citep{Abbott+2017a}, heightened interest in the coalescence of neutron stars (NSs) even outside of the field of astrophysics due to the array of fundamental physical questions that arise from the catastrophic merger \citep[see e.g.,][for comprehensive reviews]{Metzger+2017,Lattimer+2021}{}{}. A bright electromagnetic (EM) counterpart of GW170817 was a gamma-ray burst \citep[GRB; GRB 170817A;][]{Abbott+2017a}, solidifying the identification of the GW signal as that of a  binary neutron star (BNS) merger. The GRB 170817A afterglow shows evidence for a structured aspherical outflow \citep[][]{Sari+1999,Granot+2005,DuPont+2023} --- which was followed by a rise in X-ray flux at the time of writing \citep[see][for a compilation of GRB 170817A observations]{Margutti+2017,Troja+2019,Troja+2020,Troja+2022,Hajela+2022}.

Prevailing theories that explain the GW plus EM signal of GW170817 involve the BNS merger forming a short-lived remnant that eventually collapses into a black hole (BH) surrounded by an accretion disk that powers a GRB jet \citep{Murguia-Berthier+2014,Lawrence+2015}. The BH collapse is turned to as a reason to explain the production of a GRB as well as the observed lack of the $\sim\unit[10^{53}]{erg}$ of the merger remnant's rotational energy deposited into the BNS ejecta cloud \citep{Margalit+Metzger+2017}. 

In this Letter, we revisit the question of whether GW170817 may have left behind a long-lived massive NS remnant. 
We consider the case that the BNS merger creates a stable or quasi-stable massive NS that rapidly rotates to the point of centrifugally slinging matter into a ultra-relativistic equatorial outflow \citep[see e.g.,][]{Thompson+2005,Thompson+2007}\footnote{Here, ``rapid rotation'' means a spin period of $\lesssim \unit[1]{ms}$.}.\@ If the equatorial outflow is ultra-relativistic, then a sizable fraction of the NS rotational energy can be beamed into the equatorial plane. As the equatorial outflow burrows through the merger ejecta, the rapidly rotating proto-NS could (i) generate a magnetic tower that stretches along the rotation axis, thus powering a GRB jet \citep[e.g.,][]{Uzdensky+MacFadyen+2006,Uzdensky+MacFadyen+2007} and/or (ii) undergo an eventual collapse into a BH, but on a longer timescale than previously thought. 

We present the theoretical light curves of a jet-ring model in comparison with the observed data and show that late time observations can be helpful for constraining the equation of state of dense matter \citep[e.g.,][]{Tews+2018}. 

This Letter is organized as follows: Section \ref{sec:formalism} discusses the theoretical setup and dynamics of the relativistic ring produced by magneto-centrifugal slinging, Section \ref{sec:results} discusses our results in comparing our model with current observations, and Section \ref{subsec:summary} discusses our conclusions and relevance of this work. 

\section{Relativistic Ring Dynamics and Emission} \label{sec:formalism}
\begin{figure*}
    \centering
    \includegraphics[width=\textwidth]{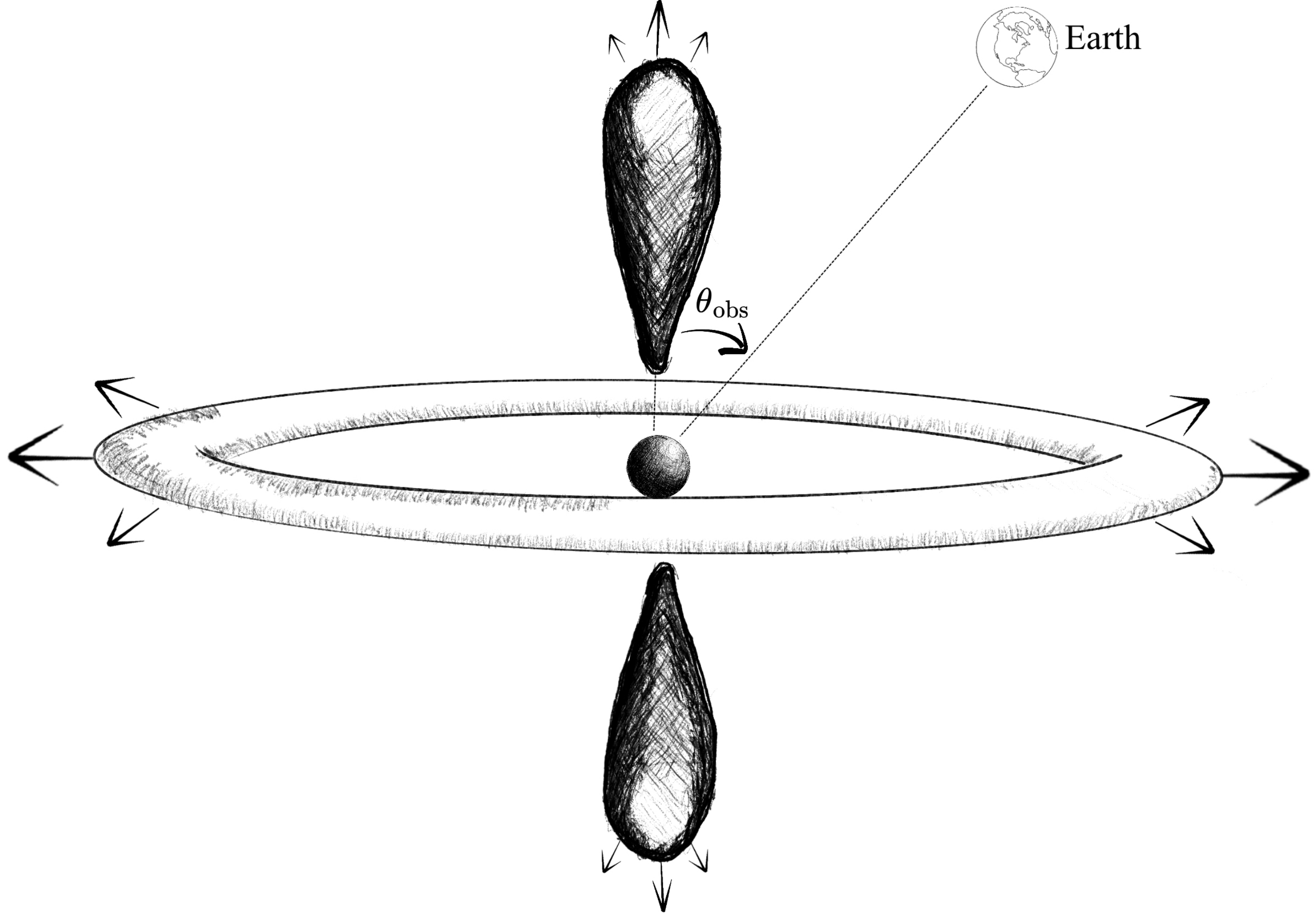}
    \caption{Illustration of the two-component emission model for GW170817 considered in this work. We assume the magnetized rotator slings a relativistically expanding, thin ring in the equatorial plane and a structured jet forms along its rotation axis. $\theta_{\rm obs}$ is taken to be the viewing angle with respect to the source's rotation axis. For brevity, the dynamical ejecta is not depicted. The arrow sizes are indicative of the relative velocity magnitudes.}
    \label{fig:cartoon}
\end{figure*}

For illustrative purposes, we use simple arguments and assumptions to derive the dynamics of a relativistic ring expelled from a magnetic rotator either through magneto-centrifugal slinging or magnetic dissipation by the equatorial current sheet. If the relativistic ring overcomes the magnetic hoop stress, it breaks out of the BNS ejecta cloud and propagates into the interstellar medium (ISM) where it will eventually sweep up mass comparable to its own and gradually open its beaming cone until visible from Earth. The BNS ejecta is assumed to produce a  collimated outflow along the rotation axis, i.e., a magnetic tower. The scenario of this structured jet-ring BNS model is illustrated in Figure \ref{fig:cartoon}.

%
%

\subsection{Toroidal outflow from a BNS merger}
The magnetic rotator's spin energy is radiated away at the rate
\begin{equation}\label{eq:edot}
    -\dot{E}_{\rm rot} = - I \omega \dot{\omega} = \frac{2}{3}\frac{\mu^2\omega^4}{c^3},
\end{equation}
where $E_{\rm rot} = M \omega^2 / 2$ is the rotational energy, $M$ is the magnetic rotator mass, $I$ is the moment of inertia, $\omega$ is the rotation frequency, $c$ is the speed of light, $\mu$ is the magnetic moment, and a dot over a variable signifies differentiation with respect to time. The rightmost equality in Equation \eqref{eq:edot} is the standard magnetic dipole radiation formula. From Equation \eqref{eq:edot}, the rotation frequency evolves as
\begin{equation}
    \omega = \omega_0 \left(1 +\frac{t}{\tau} \right)^{-1/2},
\end{equation}
where $\omega_0$ is the rotation frequency of the nascent magnetic rotator, and $\tau$ is the spin-down time. We now set $c = 1$ for the remainder of our derivations. Let us assume that outside of the light cylinder some fraction, $f$, of the rotational energy is beamed into an expanding relativistic ring with energy $(4\pi/3) \sin\theta_0r^3(\rho h\Gamma^2 - p)$, where $\theta_0$ is the initial half-opening angle of the ring's core, $\rho$ is the proper mass density, $h = 1 + \gamma p / \rho(\gamma - 1)$ is the specific enthalpy, $\gamma$ is the adiabatic index, $p$ is the pressure, $\Gamma = (1 - \beta^2)^{-1/2}$ is the Lorentz factor, and $\beta$ is the bulk velocity in units of $c$. This ring will interact with a toroidal magnetic field of strength $B_\phi$ that grows linearly with time \citep{Uzdensky+MacFadyen+2007}. If the ring is a hot relativistic fluid with adiabatic index $\gamma = 4/3$, then its pressure dominates over the rest mass energy density such that $\rho h \approx 4p$. Roughly, for this relativistic ring to overcome the magnetic field hoop stress and break out of the ejecta cloud, we require $p > B_\phi (t)^2 / 8\pi = n m_p \sigma_0 t^2/ 2 t_L^2$ where $n$ is the lab frame number density, $m_p$ is the proton mass, $\sigma_0$ is the initial magnetization, $t_L = R_{0} + \beta_{\rm c} t$ is the light crossing time across the expanding ejecta cloud, $R_{0}$ is the initial radius of the ejecta cloud, and $\beta_{\rm c}$ is the velocity of the ejecta cloud, assumed to be non-relativistic. This condition leads to an upper bound on the ``break out'' Lorentz factor of the ring,
\begin{equation}\label{eq:gamma}
    \Gamma^2(t \geq R_{0}) < \frac{f \Tilde{E}}{2\sigma_0}\left(\beta_{\rm c} + \frac{R_0}{t} \right)^{-2}\left(1 + \frac{t}{\tau} \right)^{-1},
\end{equation}
where we've neglected numerical constant terms and $\Tilde{E} \equiv 3 E_{\rm rot} / 4\pi\sin\theta_0nm_pr^3$ is the energy per baryon\footnote{$\Tilde{E}$ is related to $\eta$, the initial random Lorentz factor or the dimensionless entropy as it's often called in other texts.}. Conversely, if the ring is halted by the magnetic hoop stress it becomes baryon loaded as plasma accumulates. Eventually the rest mass energy of the ring dominates over the pressure such that $\rho h \approx nm_p / \Gamma$. These conditions might cause the ring to be redirected into a magnetic tower \citep[e.g.,][]{Uzdensky+MacFadyen+2007,Bucciantini+2012}. Immediately, one finds from Equation \eqref{eq:gamma} that for a slow-moving cavity ($\beta_c \ll R_0 /t$) and at early times ($t \ll \tau$), the breakout Lorentz factor is dominated by the ratio $(\Tilde{E} / \sigma_0)^{1/2}$ which appears to be the limiting factor that governs a successful ring breakout.
Note that Equation \eqref{eq:gamma} is an estimate and will need a direct numerical special relativistic magneto-hydrodynamics simulation to confirm its validity. 

\subsection{Dynamics of lateral spreading}
We now consider that  the relativistic ring breaks out from the BNS merger ejecta cloud and travels far into the ISM where it will eventually sweep up mass comparable to its own and begin to decelerate. At the onset of this declaration phase, we compute the dynamics of the expanding relativistic ring as it begins to spread sideways.

The energy in the blast wave is approximately
\begin{equation}\label{eq:energy}
    E \approx u^2 M = \frac{4\pi A}{3 - k}u^2r^{3 - k}\Omega,
\end{equation}
where  $[u^\mu]= \Gamma(1, \vec{\beta})$ is four-velocity, $\rho$ is the \emph{ambient} density and is taken to obey $\rho(r) = A r^{-k}$ where $A$ is the mass-loading parameter, and $\Omega$ is the blast wave solid angle.

For simplicity, we assume the solid angle of the blast wave is constant with respect to $r$ such that $M = \int_V \rho(r) r^2 dr d\Omega(r) \approx \int_\Omega d\Omega \int \rho(r) r^2dr$, unlike the more careful ``trumpet'' model computed by \citet{Rhoads+1999}. From energy conservation, we can write
\begin{equation}\label{eq:gb_derivation}
    d\ln u = - \frac{1}{2} d \ln M.
\end{equation}
Since $M \propto r^{3-k}\Omega$, the evolution of the bulk flow goes as  
\begin{equation}\label{eq:gamma_vs_rtheta}
    d \ln u = \left(\frac{k - 3}{2} \right) d\ln r - d\ln \Omega^{1/2}.
\end{equation}
The sideways spreading of the perpendicular arc length of the blast wave\footnote{Taken from the point of view of a parcel at the very edge of the shock arc.}, $r_\perp = r \theta$, can be evaluated as
\begin{equation}\label{eq:r_perp}
    dr_\perp = \left(\theta + \frac{d\theta}{d\ln r}\right)dr \approx \theta dr + \beta_\perp dt^\prime\\ 
\end{equation}
where $\beta_\perp$ is the velocity at which the shock arc opens up, $dt'$ is the infinitesimal proper time, and $\beta_\parallel$ is the velocity parallel to the bulk flow and can be taken as $\beta_\parallel \approx \beta$. By noting that the proper time transforms into the lab frame time as $dt^\prime = dt / \Gamma = dr / \Gamma \beta$ and writing $u = \Gamma \beta$, it follows that $\beta_\perp / u \approx d\theta / d\ln r$, allowing us to write Equation (\ref{eq:gamma_vs_rtheta}) as 
\begin{equation}\label{eq:}
    \frac{d \ln (u \Omega^{1/2})}{d \ln \Omega^{1/2}} = \left(\frac{k - 3}{2} \right) \frac{u}{\beta_\perp}\frac{d\theta}{d \ln \Omega^{1/2}}.
\end{equation}
By solving for the combination $u \Omega^{1/2}$, the above equation becomes separable and takes the form 
\begin{equation}
    \frac{d \ln x}{x} = \left(\frac{k - 3}{2 \beta_\perp} \right) \Omega^{-1/2}d\theta; \ x \equiv u \Omega^{1/2}.
\end{equation}
Letting $K \equiv 2(3-k)/\sqrt{\pi}\beta_\perp$, taking $\Omega = 4\pi\sin\theta_r$ with $\theta_r$ being the half-opening angle of the ring at time $t$, and assuming $\theta_r \ll 1$, we arrive at
\begin{equation}\label{eq:ur}
    u = \frac{u_0\theta_0^{1/2}  \theta_r^{-1/2}}{Ku_0\theta_0^{1/2}(\theta_r^{1/2} - \theta_0^{1/2}) + 1}.
\end{equation}
Equation (\ref{eq:ur}) poses a distinction from the jet case in that it shows algebraic spreading of the ring-like blast wave instead of exponential or logarithmic spreading of jets \citep{Granot+Piran+2012,Duffell+Laskar+2018}. 

\subsection{Semi-analytic light curves}\label{sec:flux}
The light curves are computed using the \texttt{afterglowpy} code \citep{Ryan+2020}, which assumes synchrotron radiation as the dominant emission mechanism. At its core, \texttt{afterglowpy} solves
\begin{equation}\label{eq: flux}
    F_\nu = \frac{1 + z}{4\pi d_L^2} \int_{V} \delta^2 j^\prime_\nu r^2 dr d\Omega,
\end{equation}
where $z$ is the redshift, $d_L$ is the luminosity distance, $\delta = 1 / \Gamma(1 - \vec{\beta} \cdot \hat{n})$ is the Doppler factor, $\hat{n}$ is the unit vector pointing from the observer to the source, and $j^\prime_\nu$ is the source frame emissivity. The details of the numerical scheme are explained in \citet{Ryan+2020} and therefore will not be discussed here. Instead, we list off the important parameters and their definitions in Table \ref{tab:params}. For the structured jet, we use the \texttt{GaussianCore} blast wave model provided by \texttt{afterglowpy} out of the box, and we use the GW170817 parameters given in \citet{Ryan+2020} to fit the structured jet to the early observations. To arrive at a structured ring configuration, we assume a Gaussian energy profile of the form:
\begin{equation}\label{eq: gaussian}
    E(\theta) = E_{\rm iso}e^{-(\theta_c - \theta_0)^2 / 2\theta_0^2}
\end{equation}
where $\theta_c \equiv \theta - \pi/ 2$ and $E_{\rm iso}$ is approximated as
\begin{equation}\label{eq:eiso}
    E_{\rm iso} \approx 4\pi \frac{fE_{\rm rot}}{4\pi \sin\theta_0}.
\end{equation}
The maximal rotational energy of very massive NSs can approach $E_{\rm rot} \sim \unit[10^{53}]{erg}$ \citep{Metzger+Margalit+2015}, so we use that energy as an upper bound. The fraction, $f$, is a free parameter and for comparison, we explore the respective afterglows from structured rings with values $f = 0.1$ and $f = 1$. We also note that we added the Equation \eqref{eq:ur} spreading prescription to the \texttt{afterglowpy} code to dynamically capture the off-plane afterglow rise of the toroidal outflow. We use the \texttt{FlatLambdaCDM} class in \texttt{astropy} \citep{astro13,astro18,astro22} version 6.0.0, assuming a flat $\Lambda$CDM cosmology with $\Omega_M = 0.3$ and $H_0 = \unit[70]{km~s^{-1} Mpc^{-1}}$ to convert redshift into luminosity distance. We set $k=0$ and $K$ is calculated based on Equation 9 in \citet{Ryan+2020}. The light curve parameters are highly degenerate, but we fix $n_0 \approx \unit[0.01]{cm^{-3}}$ based on upper limits inferred by \citet{Hajela+2019} and $\theta_{\rm obs} = 23^\circ$ based on constraints provided by observations \citep{Mooley+2018,Finstad+2018,Hotokezaka+2019,Ghirlanda+2019}. The data products we use were made available to us by the first author of \citet{Hajela+2022}. This concludes the ingredients needed for the analysis.

\begin{table}[t]
\tabletypesize{\scriptsize}
\centering
    \begin{tabular}{l|l}
        \hline 
        \hline
        \colhead{Parameter} & \colhead{Definition} \\ 
        \hline 
        \hline
        $E_{\rm iso}$ &  Isotropic-equivalent energy \\
        $\theta_{\rm obs}$ &  Observer angle with respect to polar axis \\
        $\theta_{0}$ &  Initial opening angle of blast wave \\
        $n_0$ &  Number density of protons \\
        $\xi_{\rm N}$ &  Fraction of electrons accelerated \\
        $z$ &  Redshift \\
        $\epsilon_{\rm e}$ &  Electron fraction of energy density \\
        $\epsilon_{\rm B}$ &  Magnetic field fraction of energy density\\
        $p$ &  Electron energy distribution index \\
         \hline
         \hline
    \end{tabular}
    \caption{Parameters relevant for computing light curves using \texttt{afterglowpy} \citep{Ryan+2020}. Note that $\theta_{\rm obs}$ is angle between the binary's angular momentum axis and the line of sight.}
    \label{tab:params}
\end{table}
\section{Results} \label{sec:results}
\begin{figure}
    \gridline{\fig{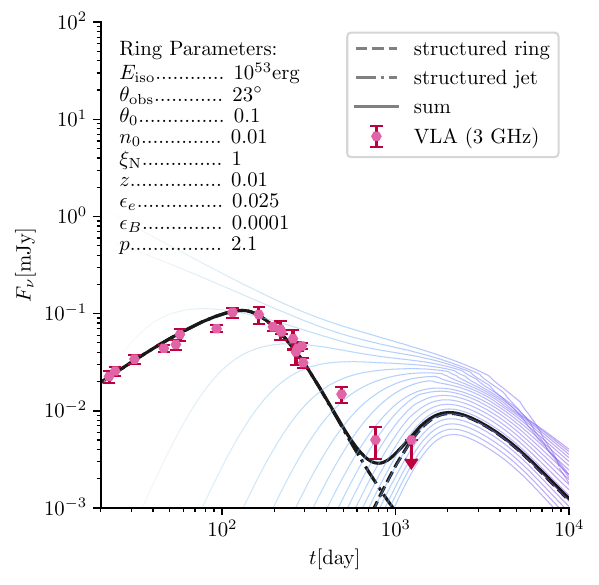}{\columnwidth}{}}
    \vspace{-3em}
    \gridline{\fig{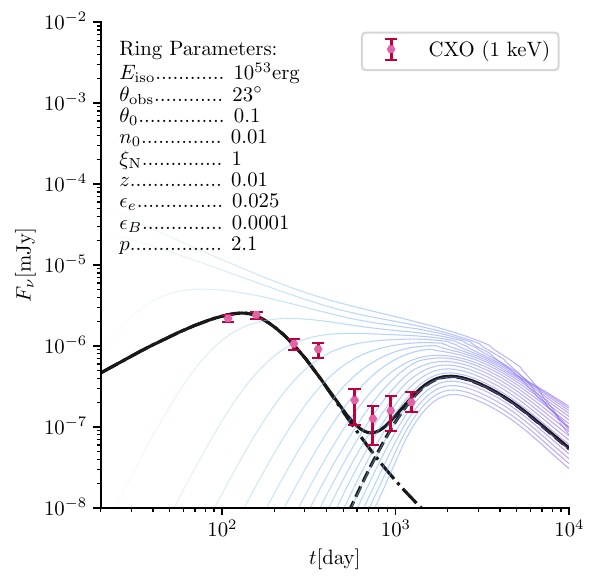}{\columnwidth}{}}
    \caption{Radio (upper) and X-ray (lower) emission profiles from GW170817 as detected by CXO and VLA instruments (pink circles, data provided by \citealt{Hajela+2022}). The black dashed curve is a representative structured ring afterglow, the black dash-dotted curve is a structured jet afterglow, and the solid black curve is the sum of both components. The fainter purple dashed curves represent a family of structured ring afterglows for various $\theta_0 \in [0.01, \pi / 2]$ and therefore $E_{\rm iso}$. All curves are computed using the \texttt{afterglowpy} code with the relevant parameters for the structured ring annotated in the upper left corner of both plots. }\label{fig:lc_52}
\end{figure}

\begin{figure}
    \gridline{\fig{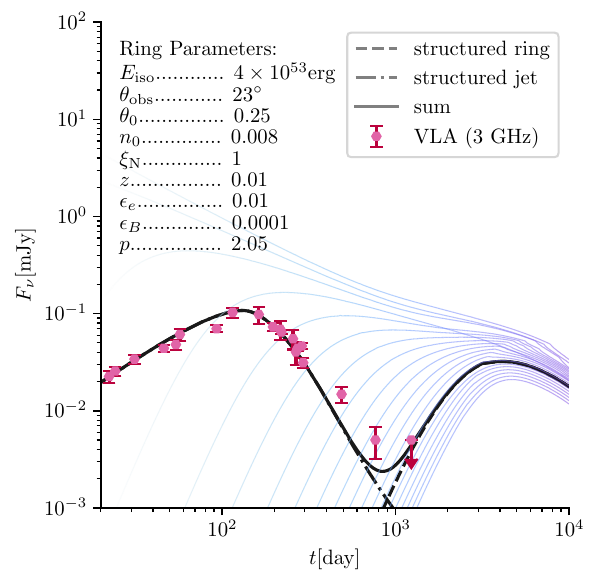}{\columnwidth}{}}
    \vspace{-3em}
    \gridline{\fig{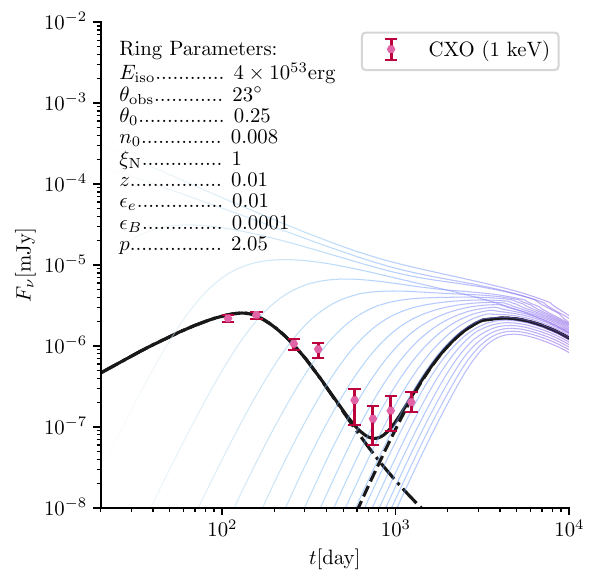}{\columnwidth}{}}
    \caption{Similar to Figure \ref{fig:lc_52} except now we assume all of the spin-down energy is beamed in the magnetic rotator's equatorial plane. The light curve parameters are also slightly modified from what is shown in Figure \ref{fig:lc_52} in order the representative dashed black curve to fit within the constraints of the observations.}\label{fig:lc_53}
\end{figure}

The results of our two-component model fitted to the GRB 170817A afterglow data are shown in Figures \ref{fig:lc_52} and \ref{fig:lc_53}. The radio data were observed using the Karl G. Jansky Very Large Array \citep[VLA; ][]{Condon+98} at 3 GHz, and the X-ray data were from the Chandra X-ray Observatory \citep[CXO;][]{Weisskpf+2000}\footnote{We manually converted the unabsorbed X-ray flux in Table 1 of \citet{Hajela+2022} to units of milli-Jansky (mJy) given the information in that table.}. 

Figure \ref{fig:lc_52} showcases the standard structured jet (black dash-dotted curve) in conjunction with our structured ring (black dashed curve) afterglow hypothesis. In this comparison, we model the ring that has $f = 0.1$ or 10\% of the magnetic rotator's rotational energy. When summed, the jet-ring model shows a decent fit with the GRB 170817A afterglow data at the time of writing. We also plot an array of hypothetical relativistic rings (purple dashed curves) with various opening angles in the range $\theta_{\rm draw} \in [\theta_0 / 2, \pi/2]$ to constrain the geometry of the outflow needed to still fit the observations\footnote{The range of opening angles also produced a range of $E_{\rm iso}$ based on Equation \ref{eq:eiso}.}. Note that the parameters in Figure \ref{fig:lc_52} are not statistically relaxed, but are only meant to demonstrate how a highly beamed, off-plane, ring-like geometry can reasonably fit the late-time observations.

As an extreme case, we also plot in Figure \ref{fig:lc_53} the structured ring light curves if all of the $\unit[10^{53}]{erg}$ rotational energy was focused into the magnetic rotator's equatorial plane (i.e., $f \sim 1$.) Under these conditions, we find that the physical parameters of the system need not change much from the previously calculated structured ring aside from varying $p$ and $n_0$ to within a few percent of their previous values and increasing the angular size of the ring from $\theta_0 = 0.1$ to $\theta_0 = 0.25$. The ring's peak brightness is almost on par with the jet's peak brightness even after 1000 days. This is due to the ring's slower spreading dynamics coupled with its energy content being larger than the jet's by a factor $\theta^{-1}$. We again plot $\theta_{\rm draw} \in [\theta_0 / 2, \pi/2]$ as purple dashed curves. Again, the parameters in Figure \ref{fig:lc_53} are not statistically relaxed, but are simply used as representative configurations of the ideal structured ring outflows that can still fit the observations given that all of the spin-down energy of the magnetic rotator is beamed in the equatorial plane. 

Equally important, although the parameters in Table \ref{tab:params} are highly degenerate, we could not find structured ring afterglows that could fit the early phase of the light curve evolution for GRB 170817A. That is to say, we find that the observations preceding the re-brightening in the afterglow light curves from GRB 170817A \emph{must} come from a structured jet-like outflow geometry due to the early decay phase being too steep to be explained by any ring-like geometry no matter what combination of parameters we use from Table \ref{tab:params}.   

\section{Discussion} \label{subsec:summary}
In this work, we calculate theoretical light curves of a jet-ring structure formed just after a binary neutron star (BNS) merger. We compare this model with the observed data from the GW170817 BNS merger event, and find that if the equatorial ring is highly beamed away from the line of sight, then its energy can be hidden until $\sim 1000$ days, which might explain the late-time excess emission observed in the afterglow \citep{Margutti+2017,Troja+2019,Troja+2020,Troja+2022,Hajela+2022}. The off-axis gamma-ray burst (GRB) jet is seen first before its afterglow begins decaying, wherein the ring's emission shines through once the ring outflow slows down enough for Earth to be inside of its beaming cone. This effect emerges as a re-brightening at late times ($\sim 1000$ days) in the afterglow. This emission differs from previous calculations of later-time ($\sim 10000$ days) bumps due to the slower dynamical ejecta of NS mergers \citep[e.g.,][]{Hotokezaka+2018}.

Based on these results, we ask whether GW170817 may have left behind a NS as opposed to collapsing into a black hole (BH). The NS collapse is usually invoked to explain the lack of $E_{\rm max} \sim \unit[10^{53}]{erg}$ rotational energy that would be deposited into the merger ejecta \citep{Metzger+Margalit+2015,Margalit+Metzger+2017}, but we consider that instead the rapidly rotating remnant beams a significant fraction of this rotational energy in the equatorial plane, and we at Earth miss the initial beamed emission and instead see the opening phase of the equatorial outflow as it decelerates. This might imply that the upper bound mass limits are higher than those previously computed for GW170817 \citep{Margalit+Metzger+2017,Shibata+2017,Ruiz+2018,Rezzolla+2018,Shibata+2019} while still falling into absolute limits set by nuclear physics \citep{Gandolfi+2012}. 

Our model is idealized in that we assume a significant fraction of the energy flux is focused into the equator where we remain agnostic about the focusing mechanism. We also do not know how long this focusing effect will last as the BNS ejecta cloud propagates outward. We compare a weakly beamed ring ($E_{\rm ring} / E_{\rm max} = 0.1$) and a highly beamed ring ($E_{\rm ring} / E_{\rm max} \sim 1$) and find that both configurations can fit the observations reasonable well, but each of these beaming modes imply a different fate for the NS. For example, if only 10\% of the $\unit[10^{53}]{erg}$ rotational energy is beamed in the equatorial plane, then the NS must eventually collapse into a BH as to not violate the electromagnetic constraints of the $\unit[10^{51}]{erg}$ kilonova (KN) ejecta from GW170817 \citep{Metzger+2017}. On the other hand, if almost all of the rotational energy \emph{is} in fact beamed into the equatorial plane, then it opens up the possibility that the remnant remains a stable NS given a stiff enough NS equation of state and/or the total mass of the binary is low \citep{Shibata+Taniguchi+2006}. 

Still, it is admittedly difficult to reconcile the fact that with the Kelvin-Helmholtz timescale being $\tau_{\rm KH} = 3GM^2/5L_\nu R \sim \unit[100M_{2.7}^2 L_{\nu 52}^{-1}R_{11}^{-1}]{s}$, the hypothetically stable NS will now be cold and the stability of hot remnants transitioning to cold slow / non- rotators is still poorly understood. In the previous estimate, $M_{2.7} = M / 2.7M_\odot$ is the dimensionless total binary mass \citep{Abbott+2019},  $L_{\nu 52} = L_\nu / \unit[10^{52}]{erg~s^{-1}}$ is the dimensionless neutrino luminosity \citep{Thompson+2004}, and $R_{11} = R / \unit[11]{km}$ is the dimensionless radius of the proto-NS inferred from general relativistic simulations \citep{Sekiguchi+2016}. While some calculations used GW170817 to weakly constrain the  maximum mass of cold NSs to $M_{\rm max} \lesssim 2.3 M_\odot$ \citep[e.g,][]{Shibata+2019}, we suggest that observations of the late-time light curve morphology might shift these upper limits towards higher values with the most extreme case being that GW170817 left behind a $2.7M_\odot$ remnant. At the most basic level, given what fraction of the magnetic rotator's energy can be feasibly hidden due to beaming in the equatorial plane such that the magnetic rotator never has to collapse into a BH to satisfy any EM constraints, our model could have significant implications on the equation of state of dense matter.

In order to compute the structure of equatorial outflows from BNS remnants we plan to perform 3D relativistic magneto-hydrodynamics simulations of a rapid magnetized rotator embedded in a BNS ejecta cloud. Previous simulations in 2D axisymmetry have shown that this sort of configuration collimates almost all of the equatorial outflow into polar jet-like outflow and therefore possesses no relativistic toroidal component \citep[e.g.,][]{Bucciantini+2007,Bucciantini+2008,Bucciantini+2012}. However, the polar focusing seen in previous works may not be a universal solution given that this focusing effect may be an artifact of the imposed axisymmetry \citep[see e.g.,][]{Porth+2014}. Furthermore, a jet-ring structure is seen in observations of the Crab pulsar \citep{Hester+1995,Hester+2002,Weisskopf+2000}, and we remain curious as to whether the canonical BNS merger event can produce a stable remnant with Crab-like morphology given that the relativistic ring successfully carries away a significant amount of the rotational energy.

\begin{acknowledgments}
The authors thank Ben Farr, Andrei Gruzinov, Aprajita Hajela, Charles Horowitz, Raffaella Margutti, Brian Metzger, Jorge Piekarewicz, Geoffrey Ryan, and Jonathan Zrake for useful discussions. We acknowledge support from NASA ATP grant 80NSSC22K0822. M.D. acknowledges a James Author Fellowship from NYU's Center for Cosmology and Particle Physics, and thanks the LSST-DA Data Science Fellowship Program, which is funded by LSST-DA, the Brinson Foundation, and the Moore Foundation; his participation in the program has benefited this work.
\end{acknowledgments}

%

\vspace{5mm}


\software{\texttt{astropy} \citep{astro13,astro18,astro22},  
          \texttt{CMasher} \citep{vanderVelden+2020},
          \texttt{afterglowpy} \citep{Ryan+2020}
          }




\bibliography{refs}{}
\bibliographystyle{aasjournal}



\end{document}